\documentclass[letterpaper, 10 pt, conference]{ieeeconf}  

\IEEEoverridecommandlockouts                              

\usepackage{graphics} 
\usepackage{epsfig} 
\usepackage{mathptmx} 
\usepackage{times} 
\usepackage{amsmath} 
\usepackage{amssymb}  
\usepackage{floatrow}
\usepackage{multirow}

\title{\LARGE \bf
Several recent developments in estimation and robust control of quantum systems*
}

\author{Daoyi Dong$^{1}$, Yuanlong Wang$^{1}$\\
{\tt\small d.dong@unsw.edu.au, yuanlong.wang.qc@gmail.com}
\thanks{*This work was supported by the Australian Research Council's Discovery Projects funding scheme under Project DP130101658.}
\thanks{$^{1}$School of Engineering and Information Technology, University of New South Wales, Canberra, ACT 2600, Australia}%
}

\begin{document}

\maketitle
\thispagestyle{empty}
\pagestyle{empty}

\begin{abstract}
This paper summarizes several recent developments in the area of estimation and robust control of quantum systems and outlines several directions for future research. Quantum state tomography via linear regression estimation and adaptive quantum state estimation are introduced and a Hamiltonian identification algorithm is outlined. Two quantum robust control approaches including sliding mode control and sampling-based learning control are illustrated.
\end{abstract}


\section{INTRODUCTION}
Quantum technology has shown powerful potential for developing future technology \cite{Nielsen and Chuang 2000}. Practical applications of quantum technology include secure quantum communication, powerful quantum computation and high-precision quantum metrology. These potential applications have attracted many mathematicians, physicists, computer scientists and control engineers to this booming field.

A fundamental task in quantum technology is to characterize the state of a quantum system and identify the parameters in the system. The estimation procedure of a static quantum state is often referred as quantum state tomography \cite{Nielsen and Chuang 2000}. For estimating a dynamical state, quantum filtering theory \cite{bouten 2007,gao 2016} has been developed, which is especially useful for addressing the measurement-based quantum feedback control problem \cite{Wiseman and Milburn 2010}. The area of identifying key parameters in quantum systems can be referred as quantum system identification \cite{burgarth 2012}. Another important problem is robustness of quantum systems in developing practical quantum technology since real quantum systems are often subject to noises, incomplete knowledge or uncertainties \cite{Dong and Petersen 2010}. In this paper, we introduce some recent developments in the areas of estimation and robust control of quantum systems. We do not intend to survey the main progress in these areas. Instead, we present several examples and methods that were recently developed by the authors and their collaborators, and aim to illustrate several classes of significant quantum estimation and control problems as well as outline open questions for future research.

This paper is organized as follows. In Section \ref{Sec2} we introduces quantum state tomography via linear regression estimation and adaptive quantum state tomography. Section \ref{sec3} presents the identification problem of quantum processes and system Hamiltonian. Section \ref{sec4} outlines two quantum robust control methods including sliding mode control and sampling-based learning control. Conclusions are presented in Section \ref{conclu}.

\section{Quantum state estimation}\label{Sec2}
\subsection{Quantum state tomography}
Quantum state tomography provides a framework to reconstruct quantum states. The state of a quantum system can be described by a density matrix $\rho$ which is a Hermitian, positive semidefinite matrix satisfying $\text{Tr}(\rho)=1$. A pure state can also be described by a unit complex vector $|\psi\rangle$ with $\rho=|\psi\rangle\langle\psi|$ \cite{Nielsen and Chuang 2000}. A mixed state is linear combination of independent pure states. To estimate a quantum state, we usually need to make measurements on many copies of the state. For quantum measurement, a set $\{P_i\}$ of positive operator valued measurement (POVM) elements is prepared (e.g., mutually unbiased basis \cite{HouOE}), where $P_i\geq 0$ and $\sum_{i}P_i=I$ with $I$ being the identity matrix. The occurrence probability of the $i$th outcome can be calculated as $p_i=\text{Tr}(\rho P_i)$ according to the Born Rule, where $\text{Tr}(A)$ returns the trace of the matrix $A$. For a given unknown quantum state, we need to design POVM measurement and develop an estimation algorithm to reconstruct the quantum state from measurement data. Various quantum state tomography methods have been developed such as maximum likelihood estimation (MLE) method \cite{paris 2004}-\cite{teo 2011}, Bayesian mean estimation approach \cite{kohout 2010,huszar 2012} and linear regression estimation (LRE) \cite{qibo}. For more details of MLE and Bayesian mean estimation, please refer to \cite{paris 2004,kohout 2010}. Here we only introduce the LRE method recently developed for quantum state tomography.

\subsection{Quantum state tomography via LRE}
In the LRE framework, the reconstruction problem of a quantum state can first be converted into a parameter-estimation problem of a linear regression model \cite{qibo}. Consider a $d$-dimensional quantum system associated with Hilbert space $\mathcal{H}$. Let $\{\Omega_{i}\}^{d^2-1}_{i=1}$ denote a  set of Hermitian operators satisfying (i) $\textmd{Tr}(\Omega_{i})=0$ and  (ii) $\textmd{Tr}(\Omega_i\Omega_j)=\delta_{ij}$, where $\delta_{ij}$ is the Kronecker function. The quantum state $\rho$ to be reconstructed can be parameterized as $$\rho=\frac{I}{d}+\sum^{d^2-1}_{i=1}\theta_i\Omega_i,$$ where $\theta_i=\textmd{Tr}(\rho\Omega_i)$. Let $\Theta=(\theta_1, \cdots, \theta_{d^2-1})^{T}$, where $T$ denotes the transpose operation. Then we parameterize the quantum measurements. Suppose a series of quantum measurements $\{E^{(j)}\}^{M}_{j=1}$ are performed. Then each operator $E^{(j)}$ can be parameterized under bases $\{\Omega_{i}\}^{d^2-1}_{i=1}$ as \cite{raqst} $$E^{(j)}=\frac{\gamma^{(j)}_{0}}{d}+\sum^{d^2-1}_{i=1}\gamma^{(j)}_{i}\Omega_i,$$
where $\gamma^{(j)}_{0}=\textmd{Tr}(E^{(j)})$ and $\gamma^{(j)}_{i}=\textmd{Tr}(E^{(j)}\Omega_i)$. Let $\Gamma^{(j)}=(\gamma^{(j)}_{1}, \cdots, \gamma^{(j)}_{d^2-1})^{T}$. When we make measurements on many identical copies of a quantum system in the state $\rho$, the probability of obtaining the result of $E^{(j)}$ can be calculated as
\begin{equation}\label{averageequation}
p(E^{(j)})=\textmd{Tr}(E^{(j)}\rho)=\frac{\gamma^{(j)}_{0}}{d}+\Theta^{T}{\Gamma^{(j)}}.
\end{equation}

Suppose that we perform $E^{(j)}$ measurements for $n^{(j)}$ times with positive results $n^{(j)}_1$ times, where $\sum_{j}^{M}n^{(j)}=N$ for the total number of copies $N$. Denote $\hat x$ the estimator of $x$. Let $\hat{p}(E^{(j)})=n^{(j)}_1/n^{(j)}$, and $e^{(j)}=\hat{p}(E^{(j)})-p(E^{(j)})$. Using (\ref{averageequation}), the following linear regression equations can be obtained for $j=1,\ \cdots,M$,
\begin{equation}\label{average2}
\hat{p}(E^{(j)})=\frac{\gamma^{(j)}_{0}}{d}+{\Gamma^{(j)}}^{T}\Theta+e^{(j)}.
\end{equation}
Now, if we obtain the solution $\Theta$, the quantum state $\rho$ can be reconstructed. We rewrite (\ref{average2}) as
\begin{equation}\label{average3}
Y_{M}=X_{M}\Theta+{{\bf{e}}_{ M}}
\end{equation}
where
$$Y_{M}=(\hat{p}(E^{(1)})-\frac{1}{d}, \cdots, \hat{p}(E^{(n)})-\frac{1}{d}, \cdots,\hat{p}(E^{(M)})-\frac{1}{d})^{T},$$ $$X_{M}=(
            \Gamma^{(1)},  \cdots,  \Gamma^{(n)},  \cdots,  \Gamma^{(M)})^{T}$$ and $${{\bf{e}}_{M}}=(e^{(1)},  \cdots,  e^{(n)},  \cdots,  e^{(M)})^{T}.$$ We aim to find an estimate $\hat{\Theta}$ such that
\begin{equation}\label{theta}
\hat{\Theta}=\underset{\hat{\Theta}}{\text{argmin}}\sum_{j=1}^{M}W^{(j)}[\hat{p}(E^{(j)})-\frac{\gamma^{(j)}_{0}}{d}-{\hat{\Theta}}^{T}{\Gamma^{(j)}}]^2,
\end{equation}
where $W_{M}=\textmd{diag}(W^{(1)},  \cdots,  W^{(n)},  \cdots,  W^{(M)})^{T}$ represent the weights of different linear regression equations. It is straightforward to obtain the least-squares solution to (\ref{theta}).
Once the solution is obtained, we can reconstruct a Hermitian matrix $\tilde{\rho}$ with $\textmd{Tr}\tilde{\rho}=1$. However, $\tilde{\rho}$ is not necessarily physical since measurement noise is unavoidable. The algorithm in \cite{smolin 2012} can be used to find a physical state $\hat\rho$ from $\tilde{\rho}$. The estimation error can be characterized using the mean squared error (MSE) $E\textmd{Tr}(\hat\rho-\rho)^2$, where $E(\cdot)$ indicates the expectation on all possible measurement outcomes.

An advantage of LRE for quantum state tomography is that its computational complexity can be characterized and the theoretical error upper bound may be obtained. In \cite{qibo}, its computational complexity $O(d^4)$ for estimating a $d$-dimensional quantum state has been proven and numerical results showed that the LRE algorithm is around 10000 times faster than the MLE approach for quantum state estimation. The LRE algorithm can be further optimized and implemented on GPU, and such improvement has demonstrated the realization of reconstructing a 14-qubit state using only 3.35 hours \cite{14bit}.

\subsection{Adaptive quantum state estimation}
\begin{figure}
\centering
\includegraphics[width=3in]{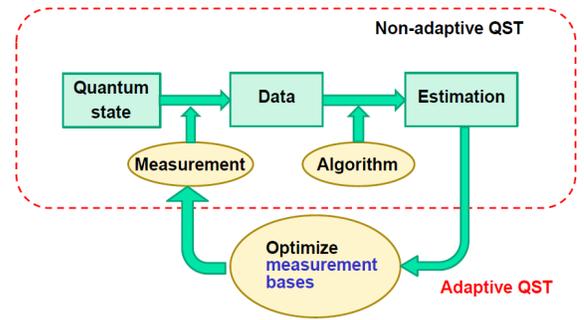}
\centering{\caption{Schematic of adaptive quantum state tomography (QST)}\label{AQST}}
\end{figure}

Another advantage of LRE is that it is suitable for developing adaptive quantum state estimation method. Adaptive protocols \cite{PRL:75:4587}-\cite{PRL:111:183601} have been proven to have the capability to improve the quantum estimation precision. In adaptive state estimation as shown in Figure \ref{AQST}, we first make measurements on part of copies and get a rough estimate of the quantum state. Then we find optimal measurement bases to make measurements on some other copies. It may involve multiple steps of adaptivity according to practical tasks.

In LRE, equation (\ref{theta}) can be recursively solved. Define
\begin{equation}\label{definitionofq}
\begin{array}{c}
Q_n=(\sum^{n}_{k=1}W^{(k)}\Gamma^{(k)}{\Gamma^{(k)}}^{T})^{-1},\\
a_n=(\frac{1}{W^{(n)}}+{{\Gamma^{(n)}}^{T}}Q_{n-1}\Gamma^{(n)})^{-1}.
\end{array}
\end{equation}
For $n=1, \cdots, M$, we have \cite{raqst}
\begin{equation}\label{recursiveofq}
Q_n=Q_{n-1}-a_nQ_{n-1}\Gamma^{(n)}{{\Gamma^{(n)}}^{T}}Q_{n-1}.
\end{equation}
$\hat{\Theta}_n$ can be recursively calculated as
\begin{equation}\label{main}
\hat{\Theta}_n=\hat{\Theta}_{n-1}+a_nQ_{n-1}\Gamma^{(n)}(\hat{p}(E^{(n)})-\frac{\gamma^{(n)}_{0}}{d}-{{\Gamma^{(n)}}^{T}}\hat{\Theta}_{n-1}).
\end{equation}

Using (\ref{recursiveofq}) and (\ref{main}), one can recursively incorporate new measurement data into historical measurement data, which provides a convenient way for adaptive estimation of $\rho$. In this sense, the LRE method is more suitable for adaptive reconstruction of quantum states due to its recursive procedure than traditional MLE or Bayesian mean method. In the LRE framework, instead of repeatedly calculating all the historical data when new data arrive, we only need to add the new data into historical information matrix and vector, which significantly reduces the calculation cost.

We need to design a criterion to optimize the measurement bases at each adaptive step. In \cite{raqst}, Qi \emph{et al.} illustrated that when the resource number $N$ becomes large enough, $E(\hat{\Theta}_n-\Theta)(\hat{\Theta}_n-\Theta)^{T}\approx Q_n$. Based on the observation, an adaptive quantum state tomography protocol has been developed in \cite{raqst}. In the first stage, one performs a standard LRE on $N_1$ copies with the standard cube measurement bases to obtain a preliminary $\hat{\Theta}$ and $Q$. In the second stage, the initial values of $Q$ in (\ref{recursiveofq}) and  $\hat{\Theta}$ in (\ref{main}) are set as $Q_0=Q$ and $\hat{\Theta}_0=\hat{\Theta}$, and then the remaining $N-N_1$ copies are utilized for multi-step adaptive estimation. If the resource number is $N_2$ in each step and $K$ steps of adaptivity are used, then $N=N_1+K\cdot N_2$.

Suppose after $s$ steps, we have $Q_{s}$ and $\hat{\Theta}_{s}$. We define $\textbf{g}_{s+1}\equiv\textmd{Tr}Q_{s}-\textmd{Tr}Q_{s+1}$. According to the experimental capability, if the candidate measurement basis set $\{E^{(s+1)}\}$ is finite, one can calculate $\textbf{g}_{s+1}$ for each candidate basis and pick up the one with the largest $\textbf{g}_{s+1}$ as the measurement basis at ($s+1$)-th step. When $\{E^{(s+1)}\}$ is an infinite set, one needs to either analytically find the optimal basis or try to obtain an approximate optimal basis. In \cite{ifac1}, an analytical optimal solution has been provided for estimating a single-qubit state. For two-qubit states, a heuristic deduction to search for the optimal measurement bases was presented in \cite{raqst}, where numerical and experimental results showed that the adaptive quantum state tomography can improve the estimation precision.

\subsection{Discussion}
Many problems remain open in the area of quantum state tomography. For example, the efficiency of the estimation algorithms may be further enhanced if there is prior knowledge on the quantum state to be reconstructed. Various variants of LRE could be developed for quantum state tomography and different adaptivity criteria can be explored for adaptive estimation of quantum states. The capability of parallel processing for quantum state estimation and the precision limit of adaptive estimation are worth further exploring. Moreover, new approaches such as coherent observers \cite{coherent-observer1,coherent-observer2} and machine learning methods may provide different angles for estimating the state of a quantum system.

\section{Quantum system identification}\label{sec3}
\subsection{Quantum Process Tomography}\label{qpt}
We illustrate the general framework of quantum process tomography described in \cite{Nielsen and Chuang 2000,chuang 1997,my 2016}. A quantum process $\varepsilon$ maps an input state $\rho_{in}$ to an output state $\rho_{out}$. In Kraus operator-sum representation \cite{Nielsen and Chuang 2000}, we have
\begin{equation}\label{kraus1}
\varepsilon(\rho_{in})=\rho_{out}=\sum_{i}A_i\rho_{in} A_i^\dagger,
\end{equation}
where $A^\dagger$ is the conjugation ($*$) and transpose ($T$) of $A$ and $\{A_i\}$ is a set of $d\times d$ matrices, with $\sum_i A_i^\dagger A_i\leq I$. We usually focus on trace-preserving operations, which means satisfying the completeness relation
\begin{equation}\label{trpreserve}
\sum_i A_i^\dagger A_i= I.
\end{equation}
By expanding $\{A_i\}$ in a fixed family of basis matrices $\{F_i\}$, we obtain $A_i=\sum_j c_{ij}F_j$, and $\varepsilon(\rho_{in})=\sum_{jk}F_j\rho_{in} F_k^\dagger x_{jk}$, with $x_{jk}=\sum_i c_{ij}c_{ik}^*$. If we take matrix $C=[c_{ij}]$, $X=[x_{ij}]$, then $X=C^TC^*$, which indicates $X$ is Hermitian and positive semidefinite. $X$ is called {\itshape process matrix} \cite{brien 2004}. $X$ and $\varepsilon$ are one-to-one correspondent. Hence, we can obtain the full characterization of $\varepsilon$ by reconstructing $X$. The completeness constraint (\ref{trpreserve}) now is $\sum_{j,k}x_{jk}F_k^{\dagger}F_j=I$.

Let $\{\rho_m\}$ be a complete basis set of the space $\mathbb C_{d\times d}$ consisting of all $d\times d$ matrices. If $\{\rho_m\}$ are linear independent, then each output can be expanded uniquely as $\varepsilon(\rho_m)=\sum_n \lambda_{mn}\rho_n$. We can establish the relationship $F_j \rho_m F_k^\dagger=\sum_nB_{mn,jk}\rho_n$. Hence, $\sum_n\sum_{jk}B_{mn,jk}\rho_n x_{jk}=\sum_n \lambda_{mn}\rho_n.$ From the linear independence of $\{\rho_m\}$, we have $\sum_{jk}B_{mn,jk} x_{jk}=\lambda_{mn}$. Let matrix $\Lambda=[\lambda_{mn}]$ and we arrange the elements $B_{mn,jk}$ into a matrix $B$:
\begin{equation}
B=\!\!
\left(\!\!\begin{array}{*{6}{c}}
B_{11,11} & B_{11,21} & \cdots & B_{11,12}  & \cdots & B_{11,d^2d^2} \\
B_{21,11} & B_{21,21} & \cdots & B_{21,12}  & \cdots & B_{21,d^2d^2} \\
\multicolumn{6}{c}{\dotfill} \\
B_{12,11} & B_{12,21} & \cdots & B_{12,12}  & \cdots & B_{12,d^2d^2} \\
B_{22,11} & B_{22,21} & \cdots & B_{22,12}  & \cdots & B_{22,d^2d^2} \\
\multicolumn{6}{c}{\dotfill} \\
B_{d^2d^2,11} & B_{d^2d^2,21} & \cdots & B_{d^2d^2,12}  & \cdots & B_{d^2d^2,d^2d^2} \\
\end{array}\!\!\right)\nonumber
\end{equation}

We define the vectorization function as $\text{vec}(A_{m\times n})= [A_{11},A_{21},...,A_{m1},A_{12},...,A_{m2},...,A_{mn}]^T$. For square $A$ we define $\text{vec}^{-1}[\text{vec}(A)]=A$. We have the relationship
\begin{equation}\label{eq3}
B\text{vec}(X)=\text{vec}(\Lambda).
\end{equation}
Here $B$ is determined once the bases $\{F_i\}$ and $\{\rho_m\}$ are chosen, and $\Lambda$ is obtained from experimental data. In practice, direct inversion or pseudo-inversion of $B$ may fail to generate a physical estimation $\hat X$ due to noise or uncertainty. A central issue in quantum process tomography is to design an algorithm to find a physical estimation $\hat X$ such that $B\text{vec}(\hat X)$ is close enough to $\text{vec}(\hat\Lambda)$. MLE \cite{jf and hradil 2001,sacchi 2001} aims to find the most likely process that can generate the current data. Bayesian mean deduction \cite{straupe,pogorelov,granade 2016} uses Bayes formula to generate a posterior probability distribution, and the expectation of this probability distribution is taken as the final estimation result. We will discuss Hamiltonian identification within the framework of quantum process tomography in the next subsection.

\subsection{Hamiltonian identification}
For a closed quantum system, its evolution can be described by
\begin{equation}
\dot{\rho}=-\frac{\text{i}}{\hbar}(\rho H-H\rho)
\end{equation}
where $\text{i}=\sqrt{-1}$, $\hbar$ is reduced Planck's constant (we set $\hbar=1$ in the following) and $H$ is the system Hamiltonian. It is clear that identifying $H$ is a fundamental task in quantum systems. Some approaches have been developed for Hamiltonian identification. For example, a Hamiltonian identification method using measurement time traces has been proposed based on classical system identification theory \cite{zhang 2014} and it has also been used to experimentally identify the Hamiltonian in spin systems \cite{zhang exp}. In \cite{wang 2015}, dynamical decoupling was employed for identifying parameters in the Hamiltonian.

Here we briefly introduce the Hamiltonian identification algorithm in \cite{my 2016}. For a closed quantum system, the state evolution can be written into $\rho_{out}=e^{-\text{i}Ht}\rho_{in}e^{\text{i}Ht}$. Compared with (\ref{kraus1}), it is clear that the unitary propagator $e^{-\text{i}Ht}$ is the only Kraus operator. Hence, $C$ is a row vector, and $X$ is of rank one. Now the semidefinite requirement is naturally satisfied. Let $X=gg^{\dagger}$ and $g=\text{vec}(G)$. We thus know $G$ must be unitary if we choose natural bases $\{|k\rangle, k=1,2,\dots\}$. Furthermore, from Theorem 2 in \cite{my 2016}, $B$ is now unitary. Denote $\hat D=\text{vec}^{-1}(B^{\dagger}\text{vec}(\hat\Lambda))$. Hence we need to find a unitary $\hat G$ to minimize $||\text{vec}(\hat G)\text{vec}(\hat G)^{\dagger}-\hat D||$ where $||\cdot||$ is the matrix Frobenius norm. This problem can be solved using a two-step optimization approach. We first find an $\hat S$ to minimize $||\text{vec}(\hat S)\text{vec}(\hat S)^{\dagger}-\hat D||$ and then find a unitary $\hat G$ to minimize $||\text{vec}(\hat G)\text{vec}(\hat G)^\dagger-\text{vec}(\hat S)\text{vec}(\hat S)^\dagger||$. Then Schur decomposition can be used to obtain the $\hat H$ through the relationship $e^{-\text{i}\hat Ht}=\hat G^T$. Further, the computational complexity $O(d^6)$ is given and an upper bound of estimation error is also established in \cite{my 2016}. Numerical results show that the Hamiltonian identification algorithm has much lower computational complexity than the approach in \cite{zhang 2014}. In \cite{Gate-identification}, it is shown that a more efficient algorithm with computational complexity $O(d^3)$ can be developed if only input pure states are used.

\subsection{Discussion}
Quantum system identification has been growing quickly in the last several years. A large number of challenging problems is waiting for exploring. For example, although several results on identifiability of quantum systems \cite{sone 2016} have been presented, the identifiability of more general quantum systems was not investigated. Adaptive approaches were only applied to the identification problems of several simple quantum systems such as estimating the Hamiltonian parameter of a two-level system \cite{wiseman ham} and more adaptive algorithms could be developed to enhance the identification precision for quantum systems. Since the computational complexity of quantum system identification algorithm usually exponentially increases with the number of qubits, it is expected to develop more efficient identification algorithms for quantum parameter identification. Other new directions for future research include the mechanism identification of physical process \cite{ShuJPCL} and the identification of quantum networks \cite{Mazzarella et al 2015,Shi et al 2014} where the network topology could be taken advantaged of.

\section{Quantum robust control}\label{sec4}
In recent years, robust control approaches have been developed to enhance the robustness performance in quantum systems. In particular, several robust control design approaches including $H^{\infty}$ control \cite{james 2008}-\cite{xcd auto}, small gain theorem and Popov method \cite{xcd tac} have been extended to linear quantum systems. More detailed results on control of linear quantum systems can refer to \cite{ian survey 2016,naoki book 2017}. In this section, we briefly introduce two robust control design methods of sliding mode control (SMC) and sampling-based learning control (SLC) for quantum systems.

\subsection{Sliding mode control}
SMC approach is a useful robust control method in classical control theory and industrial applications. However, it cannot directly be applied to quantum systems since the measurement operation usually changes the state to be measured \cite{ZhangJPB}. In \cite{0910}-\cite{1206}, a series of results on SMC of quantum systems have been presented. In particular, a quantum system with Hamiltonian uncertainty $H_{\Delta}$ has been considered and the system evolves according to Schr\"{o}dinger equation
\begin{equation}
\begin{array}{l} \text{i}|\dot{\psi}(t)\rangle
=(H_{0}+H_{\Delta}+H_{u})|\psi(t)\rangle, \\
|\psi(t=0)\rangle=|\psi_{0}\rangle ,
\end{array}
\end{equation}
where the quantum state $|\psi(t)\rangle$ corresponds to a
unit complex vector in a Hilbert space, $H_{0}$ is the free
Hamiltonian, and $H_{u}$ is control Hamiltonian. The objective is to stabilize the system in a given subspace around the target state when there exists uncertainty in the system Hamiltonian.

In order to develop an SMC approach for the robust control problem, a sliding mode $S$ is defined
as a functional of the state $|\psi\rangle$ and the Hamiltonian $H$; i.e., $S(|\psi\rangle, H)=0$. In particular, for a two-level system with the target state $|0\rangle$ (an eigenstate of $\sigma_z$ \cite{1203}), a sliding mode domain $$\mathcal{D}=\{|\psi\rangle :  |\langle
0|\psi\rangle|^{2}\geq 1-p_{0}, 0< p_{0}< 1\}$$ can be defined, where $p_{0}$ is a
given constant \cite{1203}. The definition implies
that the system's state has a probability of at most $p_{0}$ to collapse out of $\mathcal{D}$
if we make a measurement with
$\sigma_{z}$. We expect to drive and then maintain the system's state in $\mathcal{D}$.
However, $H_{\Delta}$ may take the system's state
away from $\mathcal{D}$. In \cite{1203}, a control method using the Lyapunov methodology \cite{Kuang et al 2017}, \cite{Xiaoting} and periodic projective measurements has been developed that can guarantee the desired robustness performance. The SMC idea has been used to develop a sampled-data design approach for decoherence control of a single qubit with operator errors in \cite{AAA}.

\subsection{Sampling-based learning control}\label{secadd}
Sampling-based learning control (SLC) was originally developed for control of inhomogeneous quantum ensembles \cite{Chen et al 2014} and has been used for a number of robust control problems of quantum systems \cite{Dong et al 2015}. Here we use an example to illustrate the basic idea of SLC.
Consider a finite-dimensional closed quantum system
\begin{equation}\label{ensemble control}
\begin{split}
& |\dot{\psi}(t)\rangle=-\text{i}H_{\omega,\theta}(t)|\psi(t)\rangle,~ t \in [0, T], \\
& |\psi(0)\rangle=|\psi_{0}\rangle, \\
& H_{\omega,\theta}(t)=\omega H_{0}+\theta \sum_{m=1}^{M}u_{m}(t)H_{m}, \\
& \omega \in [1-\Omega,1+\Omega], \ \ \theta \in [1-\Theta,1+\Theta],
\end{split}
\end{equation}
where $\omega$, and $\theta$ are two uncertainty parameters that can characterize inhomogeneity in quantum ensembles, uncertainties in the system Hamiltonian or fluctuations in control fields. The objective is to find a robust control field that can steer the system to a given target state $|\psi_{\text{target}}\rangle$ when uncertainties exist. We define the performance function as
$$J(u)=\vert
\langle\psi_(T)|\psi_{\text{target}}\rangle\vert^{2}.$$

The SLC method includes two steps of training and testing. In the training step, we select $N$ samples and then construct an
augmented system as follows
\begin{equation}\label{generalized-system}
\left(%
\begin{array}{c}
  |{\dot{\psi}}_{\omega_1,\theta_1}(t)\rangle \\
  |{\dot{\psi}}_{\omega_2,\theta_2}(t)\rangle \\
  \vdots \\
  |{\dot{\psi}}_{\omega_N,\theta_N}(t)\rangle \\
\end{array}%
\right)
=-\text{i}\left(%
\begin{array}{c}
  H_{\omega_1,\theta_1}(t)|\psi_{\omega_1,\theta_1}(t)\rangle \\
  H_{\omega_2,\theta_2}(t)|\psi_{\omega_2,\theta_2}(t)\rangle \\
  \vdots \\
  H_{\omega_N,\theta_N}(t)|\psi_{\omega_N,\theta_N}(t)\rangle \\
\end{array}%
\right),
\end{equation}
where
$H_{\omega_n,\theta_n}=\omega_{n} H_{0}+\theta_{n} \sum_{m}u_{m}(t)H_{m}$
with $n=1,2,\dots,N$. The performance function for the augmented
system is defined as
\begin{equation}\label{eq:cost}
J_N(u)=\frac{1}{N}\sum_{n=1}^{N}\vert \langle\psi_{\omega_n,\theta_n}(T)|\psi_{\text{target}}\rangle\vert^{2}.
\end{equation} The goal of the training step is to find a control field $u^*$
that maximizes the performance function in \eqref{eq:cost}. The gradient flow algorithm has been developed for achieving this goal for several classes of quantum robust control problems.
Then in the testing step we apply the optimal
control $u^{*}$ obtained in the training step to additional
samples to evaluate the control performance of each sample.
If the performance for all the
tested samples is satisfactory, we accept the designed control
law. Otherwise, we need to improve the algorithm to achieve acceptable performance. The SLC method has been successfully applied to many quantum  control tasks including control and classification of inhomogeneous quantum ensembles \cite{Chen et al 2014}, \cite{quantum classification}, robust control of quantum superconducting systems \cite{Dong 2015SR}, learning robust pulses for generating universal quantum gates \cite{Dong 2016SR} and synchronizing collision of molecules with shaped laser pulses \cite{Zhang 2016RSC}. Other machine learning algorithms \cite{DONG-QRL}, \cite{Dong-QRL2008}, \cite{Chen and Dong et al 2014} can be easily integrated into the SLC method. Recently, the SLC method has been integrated into an improved differential evolution algorithm for control fragmentation of halomethane
molecules CH$_{2}$BrI using femtosecond laser pulses \cite{DE2017}.

\section{Conclusion}\label{conclu}
We introduced some recent progress in the areas of quantum state estimation, quantum Hamiltonian identification and quantum robust control. A large number of open questions remain in these emerging areas, and systems control theory may make more contributions to address these challenging issues by integrating it with the unique characteristics of quantum systems.



\end{document}